\documentclass[11pt]{article}

\input epsf.sty

\usepackage{graphicx,amsmath,amssymb}
\def\lromn#1{\uppercase\expandafter{\romannumeral#1}}

\begin{document}

\begin{flushright}
\today \\
\end{flushright}

\begin{center}
\begin{Large}
{\bf Determination of CP violation parameter 
using neutrino pair beam
}
\end{Large}

\vspace{2cm}
M.~Yoshimura and N. Sasao$^{\dagger}$

\vspace{0.5cm}
Center of Quantum Universe, Faculty of
Science, Okayama University \\
Tsushima-naka 3-1-1 Kita-ku Okayama
700-8530 Japan

\vspace{0.2cm}
$^{\dagger}$
Research Core for Extreme Quantum World,
Okayama University \\
Tsushima-naka 3-1-1 Kita-ku Okayama
700-8530 Japan \\

\end{center}

\vspace{5cm}

\begin{center}
\begin{Large}
{\bf ABSTRACT}
\end{Large}
\end{center}

Neutrino oscillation experiments under
neutrino pair beam from circulating excited heavy
ions are studied.
It is found that detection of double weak events
has a good sensitivity to measure
CP violating parameter and distinguish mass hierarchy
patterns in short baseline experiments in which the
earth-induced matter effect is minimized.

\vspace{4cm}
%\pacs{ 
PACS numbers
\hspace{0.5cm} 
13.15.+g, %% Neutrino interactions  
14.60.Pq, %% Neutrino mass and mixing 

Keywords
\hspace{0.5cm} 
CP violation, 
neutrino oscillation,
CP-even neutrino  beam, 
heavy ion synchrotron, 
earth matter effect

\newpage

\section
{\bf Introduction}

As  is pointed out in our previous paper \cite{pair beam},
when excited ions with a high coherence are
circulated, neutrino pair emission rates become large with neutrino
energies extending  to the GeV region.
Produced neutrino beam is a coherent mixture of
all pairs of neutrinos, $\nu_e\bar{\nu}_e\,, \nu_{\mu} \bar{\nu}_{\mu}\,, \nu_{\tau} \bar{\nu}_{\tau}$.
This gives a CP-even neutrino beam,
providing an ideal setting to
test fundamental symmetries of particle physics \cite{cpt theorem and sterile},
in particular, to  measure the CP violating (CPV) phase in the neutrino sector \cite{cp violation}.

In the present work sequel to the previous one,
we investigate observable quantities at detection sites
away from  heavy ion synchrotron, including
the location of  the facility.

Our main physics objectives are

(1) CPV $\delta$ phase measurement
(excluding the ones intrinsic to the Majorana neutrino),

(2)  NH vs IH distinction.

We shall demonstrate that double neutrino detection is 
necessary to achieve these objectives.
Furthermore, in order to avoid possible contamination of
earth-induced effects that mimic CPV parameter dependence,
it is wise to conduct oscillation experiments at
a short baseline.
Our results show that a location within $\sim$ 50 km away from
the synchrotron can do an excellent job.

The rest of this paper is organized as follows.
In the second section we explain special features of
neutrino oscillation experiments under CP-even neutrino
pair beam after a brief summary of neutrino
pair emission at synchrotron.
The conclusion on experimental means is that
one should measure double weak events at
detector for CPV parameter determination.
The neutrino pair beam is found insensitive to
CPV phases intrinsic to the Majorana neutrino.
In Section 3 we discuss short baseline experiments
in which the earth-induced matter effect is neglected.
We demonstrate that both CP-even
and CP-odd  quantities can provide a sensitive
measurement of CPV parameter with high precision. 
Distinction of normal and inverted hjerarchical mass patterns
is shown to be possible in short baseline experiments of the pair beam.
In Section 4 the earth matter effect is discussed and
shown to give large influence on determination of CPV parameter.

Throughout this work we use the natural unit of $\hbar = c = 1$.

\vspace{0.5cm} 
\section
{\bf Neutrino experiments under coherent pair beam}

We consider measurements of neutrinos at a distance $L$
under the coherent neutrino pair beam of all mixtures of
$\nu_a \bar{\nu}_a, a = e, \mu, \tau$ produced at a heavy ion synchrotron.
It is important to calculate detection rates by treating
the whole event quantum mechanically, since
produced neutrino pairs are not detected at the synchrotron site.

We shall first recapitulate main features of the coherent
neutrino pair beam proposed in our previous paper \cite{pair beam}.
The neutrino pairs are  produced from excited heavy ions
of a boost factor $\gamma$  circulating in a ring of radius $\rho$.
Its production rates are enormous, given by
\begin{eqnarray}
&&
\Gamma_{2\nu} \sim 3.1 \times 10^{21} {\rm Hz}
\frac{ N |\rho_{eg}|^2 }{10^8} (\frac{\rho}{ 4 {\rm km}})^{1/2} 
(\frac{\gamma}{10^4})^4 (\frac{\epsilon_{eg}}{50 {\rm keV}})^{11/2}
\,,
\label {pair production rate}
\end{eqnarray}
ignoring the ion spin factor of order unity.
The ionic coherence $\rho_{eg}$ between the excited and the ground levels
of spacing $\epsilon_{eg}$ is required to be substantial, and we assumed
in this estimate a number $10^8$ when it is multiplied by
the total available ion number $N$.
Relation between the pair production amplitude $ {\cal P}_{\bar{b}b} (1,2)$
of a neutrino pair $\bar{\nu}_b \nu_b\,, b = e,\mu,\tau$ with kinematical variables
collectively denoted by 1,2 (angles measured from the ion tangential direction),  and its
rate $R_{\bar{b}b} (1,2)$ is \cite{pair beam} 
\begin{eqnarray}
&&
R_{\bar{b}b} (1,2)= 2 \frac{ | {\cal P}_{\bar{b}b} (1,2)|^2 }{T} 
\,, \hspace{0.5cm}
T = \frac{\sqrt{\pi}}{2^{1/4}} \sqrt{\rho} F^{-1/4}
\,,
\\ &&
F = (E_1 + E_2) (\frac{\epsilon_{eg}} {\gamma} - \frac{E_1+E_2}{2\gamma^2})
- \frac{1}{2}(E_1^2 \psi_1^2 + E_2^2 \psi_2^2) -\frac{E_1E_2}{2}(\theta_1- \theta_2)^2
- \frac{\epsilon_{eg}}{2\gamma} ( E_1\theta_1^2 +  E_2\theta_2^2)
\,.
\label {differential spectral rate h2}
\end{eqnarray}
$T/2$ typically of order 10 ps is the effective time of neutrino pair emission, which
is more precisely a function of neutrino energies and their emission angles.
Vertical emission angles $\psi_i, i = 1,2$ and the opening angle of a pair,
%with its cosine given by
$\sin^{-1} (\cos \psi_1 \cos \psi_2 \cos (\theta_1 - \theta_2))$, 
are both limited by the boost factor $1/\gamma$.
These angles are of order $100 \,\mu$radian $10^4/\gamma$.

The probability amplitude of the entire process consists of three parts:
the production, the propagation, and the detection due to
charged current (CC) interaction, each
to be multiplied at the amplitude level.
Thus, one may write the amplitude for double neutrino quasi-elastic scattering
(with $J$ the nucleon weak current) as
\begin{eqnarray}
&&
\sum_b
(\frac{G_F}{\sqrt{2}})^2 \bar{\nu}_a \gamma_{\alpha} ( 1- \gamma_5) l_a J^{\alpha}
\bar{l}_c \gamma_{\beta} ( 1- \gamma_5) \nu_c  (J^{\beta})^{\dagger}
\langle \bar{a} | e^{-iH L} | \bar{b} \rangle  \langle c | e^{-iH L} | b \rangle 
{\cal P}_{\bar{b}b}(1,2)
\,,
\end{eqnarray}
where  $H$ is the hamiltonian for propagation including
earth-induced matter effect \cite{wolfenstein}, \cite{barger etc}, \cite{xing}, which is in the flavor basis
\begin{eqnarray}
&&
H = U \left(
\begin{array}{ccc}
\frac{m_1^2} {2E} & 0  & 0\\
0 & \frac{m_2^2} {2E}  & 0\\
0 & 0  & \frac{m_3^2} {2E}
\end{array}
\right) U^{\dagger}
\mp \sqrt{2} G_F n_e \left(
\begin{array}{ccc}
1 & 0  & 0\\
0 &  0 & 0\\
0 & 0  &0
\end{array}
\right)
\,,
\end{eqnarray}
with $U = (U_{ai}), a = e, \mu, \tau, i = 1,2,3$ the neutrino mixing matrix.
The sign $\mp$ refers to neutrino (-) and anti-neutrino (+).

Let $V$ and $\bar{V}$ are  unitary $3\times 3$ matrices that diagonalize
the hamiltonian $H$ for neutrino and $\bar{H}$ for anti-neutrino, including the earth matter effect.
We shall denote three eigenvalues by $\lambda_i$ for neutrinos,
and $\bar{\lambda}_i$ for anti-neutrinos.
The propagation amplitude is then
\begin{eqnarray}
&&
\langle c | e^{-iH L} | b \rangle = \sum_i V_{ci}^* V_{bi} e^{-i\lambda_i L}
\,, \hspace{0.5cm}
\langle \bar{a} | e^{-iH L} | \bar{b} \rangle = \sum_i \bar{V}_{ai}^* \bar{V}_{bi} e^{-i \bar{\lambda}_i L}
\,,
\\ &&
\sum_b
\langle \bar{a} | e^{-iH L} | \bar{b} \rangle  \langle c | e^{-iH L} | b \rangle c_b =
\frac{1}{2}\sum_{ij} V_{ci}^* \bar{V}_{a j}^* \xi_{ij} e(\bar{\lambda}_j, \lambda_i)
\,, \hspace{0.5cm}
(c_b )=  \frac{1}{2} (1,-1,-1) 
\,,
\\ &&
\xi_{ij} =  
\bar{V}_{ej} V_{ei} - \bar{V}_{\mu j} V_{\mu i} - \bar{V}_{\tau j} V_{\tau i}
\,, \hspace{0.5cm}
e(\bar{\lambda}_j, \lambda_i)  = \exp[-i L( \lambda_i + \bar{\lambda}_j)]
\,.
\end{eqnarray}
The factor $c_b$ arises from the production amplitude ${\cal P}_{\bar{b}b}(1,2)$.
The precise relation between neutrino and anti-neutrino eigenvalue problem is given by
\begin{eqnarray}
&&
\bar{\lambda}(G_F) = \lambda(-G_F)
\,, \hspace{0.5cm}
\bar{V}_{ai}^* (G_F) = V_{ai} (-G_F)
\,.
\end{eqnarray}

An important question of the Majorana CPV phase (MP) dependence of
the neutrino propagation amplitude $\langle a| e^{-iHL } | b \rangle$
and its anti-neutrino counterpart is worked out
as follows, using the parametrization \cite{mixing parameters}.
First, the eigenvalue equation
${\rm det} (\lambda - H) = 0$,
when explicitly written out,
indicates that $\lambda_i, \bar{\lambda}_i$ 
are independent of MP, $\alpha, \beta$.
Define MP-independent mixing matrix by $\tilde{U} = U P^{\dagger}\,,
P = (1, e^{i\alpha}, e^{i\beta})$.
The hamiltonian in the mass eigen-state basis
$ U^{\dagger} H U$ has a simple MP phase dependence
$P^{\dagger} \tilde{H} P$, $\tilde{H}$ being MP-independent.
Diagonalization of $\tilde{H}$ can be done,
$\tilde{H} = \tilde{V}^{\dagger} H_D \tilde{V}$ by
MP-independent matrix $\tilde{V}$.
The unitary matrix $V$ for $H$ diagonalization is
then MP-independent, since
$V = \tilde{V} P U^{\dagger} = \tilde{V}  P P^{\dagger} \tilde{U} ^{\dagger} = \tilde{V} \tilde{U} ^{\dagger} $.
This proves that $\langle a| e^{-iHL } | b \rangle$
is MP-independent.

More general formulas relating these to re-phasing invariant quantities
are given in \cite{xing}.

We now discuss prospects of single neutrino events
in which one of pair neutrinos go undetected.
The rate of neutrino $\nu_c$ undetected (and $\bar{\nu}_{\mu}$
detected) contains the squared propagation factor,
\begin{eqnarray}
&&
\sum _c | \sum_{ij} V_{ci}^* \bar{V}_{\mu j}^* \xi_{ij} e(\bar{\lambda}_j, \lambda_i) |^2 
= \sum_{ijkl} \sum_c V_{ci}^*  V_{ck} \bar{V}_{\mu j}  \bar{V}_{\mu l}^*
\xi_{ij} \xi_{kl}^* e(\bar{\lambda}_j, \lambda_i) e^*(\bar{\lambda}_l, \lambda_k)
\nonumber \\ &&
 = \sum_{jl} \bar{V}_{\mu j}  \bar{V}_{\mu l}^* e(\bar{\lambda}_j, \lambda_i) e^*(\bar{\lambda}_l, \lambda_i)
\sum_i \xi_{ij} \xi_{il}^*
= \sum_{jl} \bar{V}_{\mu j}  \bar{V}_{\mu l}^*  e(\bar{\lambda}_j, \lambda_i) 
e^*(\bar{\lambda}_l, \lambda_i) \delta_{jl} = 1
\,,
\end{eqnarray}
since $e(\bar{\lambda}_j, \lambda_i) e^*(\bar{\lambda}_l, \lambda_i)$ is $i-$independent.
This result relies on the unitarity of mixing matrices alone.

A conclusion drawn from this is that
when only one of neutrinos in the pair is detected
and its partners are undetected,
oscillation patterns disappear, hence there is no way to measure
CPV parameter.

We thus consider double events in what follows.
The single event may be used to monitor the coherence $\rho_{eg}$
which might be otherwise not easy to measure.

\vspace{0.5cm}
\section
 {\bf Neutrino interaction away from synchrotron: short baseline experiments}

An important question that arises in the coherent neutrino pair beam is whether 
the coherence of two neutrino of the pair present at the production site  is
maintained or not.
The degree of phase de-coherence increases with travel distance of the
neutrino pair.
The most important  de-coherence interaction
is forward scattering  caused by atomic electrons \cite{wolfenstein} when
the neutrino passes through the earth.
This introduces a phase of order,
\begin{eqnarray}
&&
\sqrt{2} G_F n_e L \sim 1 \times \frac{n_e}{1.4 \times 6 \times 10^{23}{\rm cm}^{-3}}
\frac{L}{1860 {\rm km}}
\,,
\end{eqnarray}
(for the simple earth-model made of  pure SiO$_2$ of mass density 2.8 g cm$^{-3}$) and 
its fluctuating component destroys the phase coherence necessary
to apply the idea of the coherent pair beam.
Thus, we may divide the nature of the pair beam into coherent and incoherent
regimes, its boundary being roughly estimated at of order 2000 km.

Double event detection probability of quasi-elastic scattering (QES) 
producing two charged leptons $\mu^+$ and $ c (= e, \mu)$
is determined from the product of three factors,
\begin{eqnarray}
&&
 | \sum_{ij} V_{ci}^* \bar{V}_{\mu j}^* \xi_{ij} e(\bar{\lambda}_j, \lambda_i) |^2 
\frac{d^4 \Gamma}{dE_1 dE_2 d\Omega_1 d\Omega_2 }
\frac{d^2 \sigma}{d E_+ d\sin \psi_+}\frac{d^2 \sigma}{d E_- d\sin \psi_-}
\,.
\label {double event rate}
\end{eqnarray}
$\mp $  corresponds to neutrino ($\nu_c$) and anti-neutrino ($\bar{\nu}_{\mu}$) events.
One can assume for parent neutrino pairs that $i=1$ for $\bar{\nu}_a, a = \mu$
and  $i=2$ for $\nu_c$.

The probability of double detection is roughly estimated as follows.
First, the detection probability of a single neutrino event is
 estimated by the factor $\sigma n_N l$
where $n_N$ is the nucleon number density and $l$
is the detector's size along the neutrino beam.
The cross section is of order $10^{-39} \sim 10^{-38}$cm$^2$
for a 1 GeV neutrino, which gives
$\sigma n_N l \sim 10^{-11} \sim 10^{-10}$ for 
a single detection of  weak process with $\sim$100 m detector size.
The double detection probability is then, with
a perfect acceptance, $10^{-22} \sim 10^{-21}$
for a 100 kt class of detectors, which gives the double rate
of order $10 \sim 100$ mHz, using the formula (\ref {pair production rate}).
This rate is not too small.
Actual experimental design, which we do not discuss in this paper, must take into account detailed rate calculation 
including detector geometry, location etc. as well as possible backgrounds.

In this section we shall consider short baseline oscillation experiments
ignoring earth-induced matter effects, hence
the propagation factors of eq.(\ref{double event rate}) becomes
\begin{eqnarray}
&&
P_{\bar{a} c} = | \sum_{ij} U_{ci}^* U_{aj} \xi_{ij} e^{-i L(m_j^2/E_1 + m_i^2/E_2)} |^2 
\,, \hspace{0.5cm}
\xi_{ij} =  
U_{ei} U_{ej}^* - U_{\mu i} U_{\mu j}^* -  U_{\tau i} U_{\tau j}^*
\,.
\label {coherent propagation formula}
\end{eqnarray}
Striking unique feature of this formula for $P_{\bar{a} c}$ is that 
phase modulations caused by oscillation effects
emerge at relatively short distances of measurement sites,
because the oscillation phase 
given by $ L ( m_j^2/E_1 +  m_i^2/E_2)$ in eq.(\ref {coherent propagation formula})
contains two terms of $L/E_i, i= 1,2$ with two neutrino energies $E_i$
constrained only by $E_1 + E_2 \leq E_m, E_m = 2 \epsilon_{eg} \gamma$.
When $E_2 \ll E_1$, even a large $E_1$ case 
has contribution of large phases from larger values of $L/E_2$.

Interesting oscillation patterns begin to appear 
already at distance 10 km away, essentially at the synchrotron site.
CPV parameter can be determined  by measurements of both
CP-even and CP-odd quantities.
A typical CP-odd quantity, CPV asymmetry  for the rate difference
of $\bar{\nu}_a \nu_c$ and  $\bar{\nu}_c \nu_a$ events, is defined by
the ratio of rate difference to the rate sum;
\begin{eqnarray}
&&
A(\delta) = \frac{d\Gamma(\delta: G_F)  - d\Gamma( -\delta: -G_F)  }
{d\Gamma( \delta: G_F)  + d\Gamma(-\delta: -G_F) }
\,.
\label {cpv asymmetry def}
\end{eqnarray}
The change $G_F \rightarrow - G_F$ is necessary when the earth matter effect is included
in the next section.

\begin{figure*}[htbp]
 \begin{center}
 \epsfxsize=0.6\textwidth
 \centerline{\epsfbox{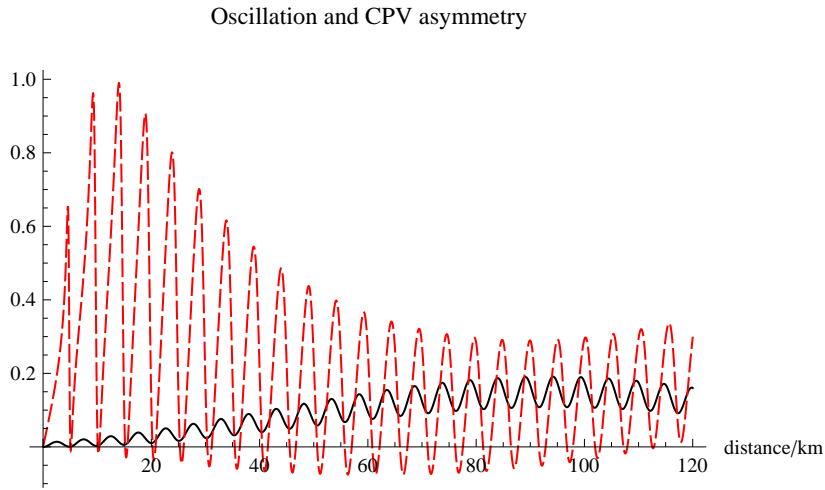}} \hspace*{\fill}
   \caption{Oscillation pattern given by 
$P_{\bar{\mu}e }$ of eq.(\ref{coherent propagation formula})
 (in solid black) and asymmetry (in dashed red) 
at various distances for
$\bar{\nu}_{\mu} \nu_e$ CC double  events.
$\delta = \pi/4, E_{\bar{\nu}_{\mu}} = 500 {\rm MeV}, E_{\nu_e} = 5$MeV.
}
   \label {oscillation and asymmetry em}
 \end{center} 
\end{figure*}

\begin{figure*}[htbp]
 \begin{center}
 \epsfxsize=0.6\textwidth
 \centerline{\epsfbox{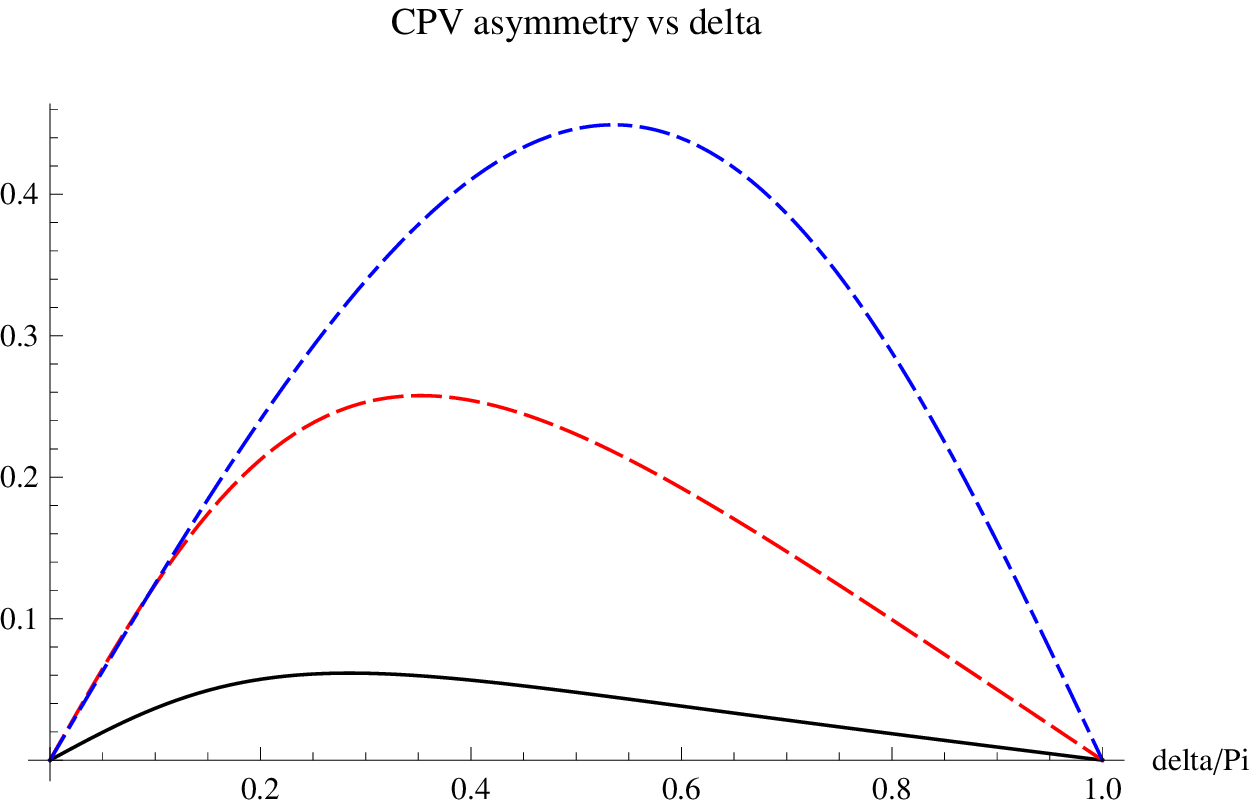}} \hspace*{\fill}
   \caption{Asymmetry vs CPV $\delta$ for
$\bar{\nu}_{\mu} \nu_e$ CC double  events.
$E_{\bar{\nu}_{\mu}} = 500 {\rm MeV},E_{\nu_e} = 5 $MeV
at 10 km away in solid black, 50km in dashed red, and 100km in dash-dotted blue.
NH of smallest mass zero is assumed.
}
   \label {asymmetry vs delta}
 \end{center} 
\end{figure*}

Computed oscillation patterns given by
$P_{\mu^+ e^-}$ and asymmetries calculated from this quantity 
are illustrated in Fig(\ref{oscillation and asymmetry em})
$\sim$ Fig(\ref{asymmetry vs delta}).
For these figures we assume the normal hierarchical (NH) mass pattern
of the vanishing smallest neutrino mass, using
oscillation parameters as given in \cite{mixing parameters}.
In these computations reaction cross sections are not multiplied, and
we did cut off the lowest neutrino energies at much larger than
$m_e$ and $m_{\mu}$ to ignore threshold effects of  charged current (CC) interactions.

These results indicate expected behaviors of oscillation patterns and
CPV asymmetry in short baseline experiments limited to distances shorter than $\sim$ 100 km:

(1) CPV asymmetry is large and of order unity near the synchrotron site, while
CP-even rates of $\bar{\nu}_{\mu} \nu_e +
\bar{\nu}_e \nu_{\mu}$ becomes  larger further away from the site.

(2) $\bar{\nu}_{\mu} \nu_e$  double events, in particular their asymmetric events
of $E_{\mu} \gg E_e$,  have larger CPV asymmetry than
$\bar{\nu}_{\mu} \nu_{\mu}$ events.

\begin{figure*}[htbp]
 \begin{center}
 \epsfxsize=0.6\textwidth
 \centerline{\epsfbox{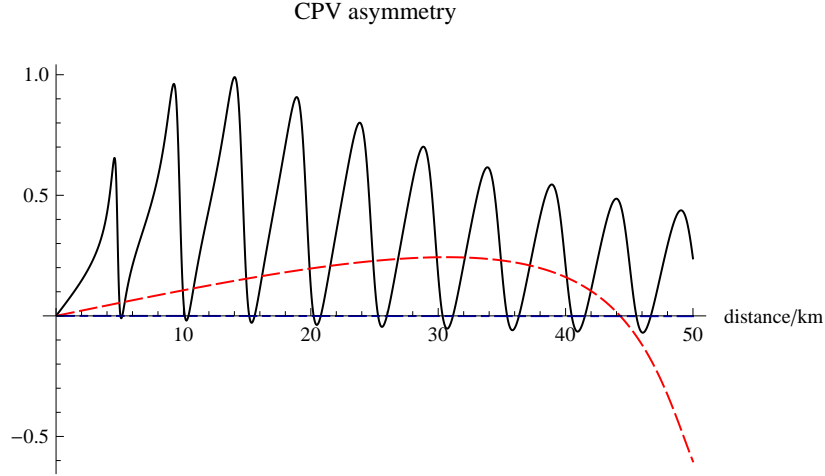}} \hspace*{\fill}
   \caption{Asymmetry at various distances for
$\bar{\nu}_{\mu} \nu_e$ CC double  events.
$\delta = \pi/4, E_{\bar{\nu}_{\mu}} = 500 {\rm MeV}$ and $E_{\nu_e} = 5$MeV
in solid black, 50 MeV in dashed red, and 500 MeV in dash-dotted blue
(much smaller than the other two cases).
NH of smallest mass zero is assumed.
}
   \label {asymmetry vs distance}
 \end{center} 
\end{figure*}

\begin{figure*}[htbp]
 \begin{center}
 \epsfxsize=0.6\textwidth
 \centerline{\epsfbox{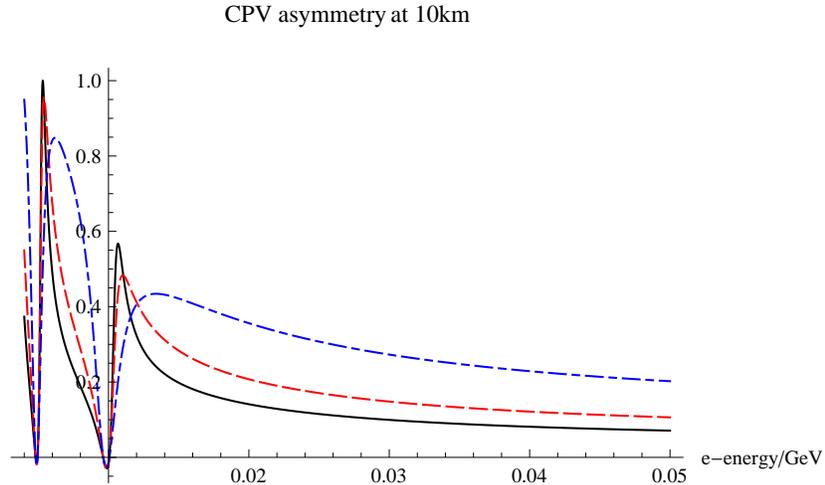}} \hspace*{\fill}
   \caption{Asymmetry vs electron neutrino energy for
$\bar{\nu}_{\mu} \nu_e$ CC double  events.
$ E_{\bar{\nu}_{\mu}} = 500 {\rm MeV}$ and
$\delta = \pi/6 $
in solid black, $\pi/4$ in dashed red, and $\pi/2$ in dash-dotted blue.
NH of smallest mass zero is assumed.
}
   \label {asymmetry vs e-energy}
 \end{center} 
\end{figure*}

\begin{figure*}[htbp]
 \begin{center}
 \epsfxsize=0.6\textwidth
 \centerline{\epsfbox{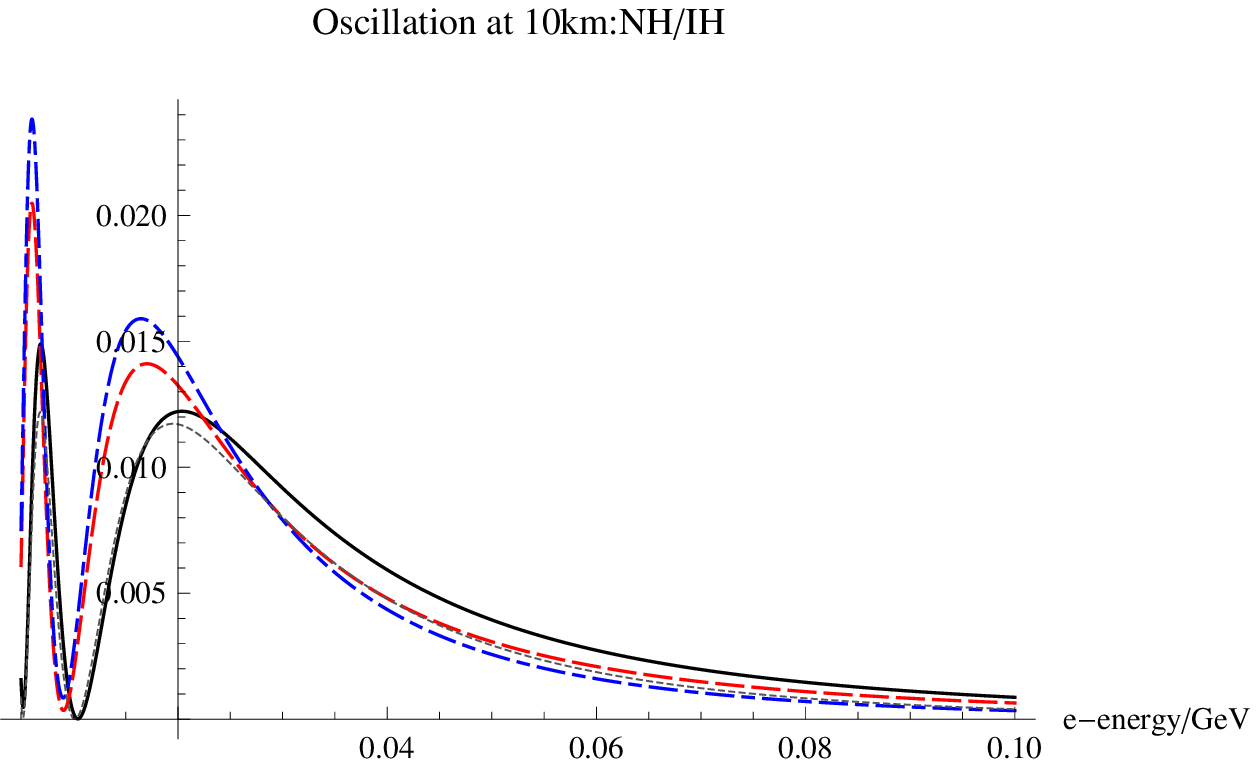}} \hspace*{\fill}
   \caption{NH vs IH distinction at 10 km away from
the synchrotron, given by asymmetric energy
combinations:
$P_{\bar{\mu}e }$ is plotted for  $E_{\bar{\nu}_{\mu}} = 500, 200$MeV, fixed and variable $ E_{\nu_e}$.
 NH in blacks, 500 MeV in solid and 200 MeV in dotted lines, and
IH in colored, 500MeV,
in dashed red  and 200 MeV in dash-dotted blue.
$\delta = 0$.
}
   \label {nh vs ih}
 \end{center} 
\end{figure*}

Finally, we show differences of normal hierarchical mass pattern (NH) 
and inverted hierarchical pattern (IH) in Fig(\ref {nh vs ih}).
There is no problem of distinction between these two cases
in short baseline experiments of neutrino pair beam.

Requirement for an effective detector to measure these quantities
is a good separation of $\mu^{\pm}$ charges, and
a good position detection of $e^{\pm}$ showers.

\vspace{0.5cm}
\section
 {\bf Comparison with long baseline experiments}

It is well known that the earth matter effect fakes CPV measurement,
and we shall examine this issue in experiments under the coherent neutrino pair beam.
In our analysis we use correction to mass eigenvalues and
eigen-vectors due to earth matter effect:
\begin{eqnarray}
&&
\lambda_1 \sim  \frac{m_1^2}{2E} - \sqrt{2} G_F n_e |U_{e1}|^2 
\,, \hspace{0.5cm}
\lambda_2 \sim   \frac{m_2^2}{2E} - \sqrt{2} G_F n_e |U_{e2}|^2 
\,, \hspace{0.5cm}
\lambda_3 \sim  \frac{m_3^2}{2E} - \sqrt{2} G_F n_e |U_{e3}|^2
\,,
\\ &&
\langle \lambda_1| a \rangle \sim
\left(
U_{a1} +2 \sqrt{2}G_F n_e E U_{e1}
(\frac{ U_{a2} U_{e2}^* }{\delta m_{21}^2} + \frac{ U_{a3} U_{e3}^* }{\delta m_{31}^2} )
\right)
\,,
\\ &&
\langle \lambda_2| a \rangle \sim
\left(
U_{a2} +2 \sqrt{2}G_F n_e E U_{e2} 
( \frac{ U_{a1}U_{e1}^*  }{\delta m_{12}^2} + \frac{ U_{a3} U_{e3}^* }{\delta m_{32}^2} )
\right)
\,,
\\ &&
\langle \lambda_3| a \rangle \sim
\left(
U_{a3} +  2 \sqrt{2}G_F n_e E U_{e3}
( \frac{ U_{a1}U_{e1}^*  }{\delta m_{13}^2} + \frac{ U_{a2} U_{e2}^* }{\delta m_{23}^2} )
\right)
\,.
\end{eqnarray}
Note the trivial relation $\delta m_{ij}^2 = - \delta m_{ji}^2 $.

\begin{figure*}[htbp]
 \begin{center}
 \epsfxsize=0.6\textwidth
 \centerline{\epsfbox{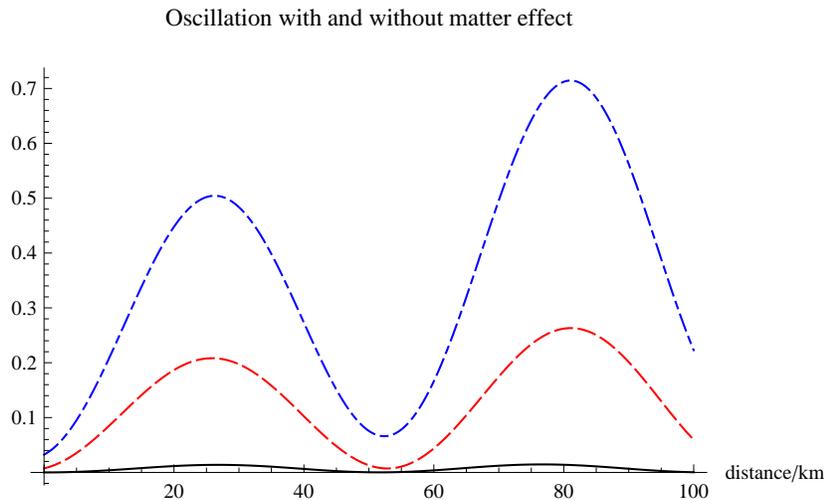}} \hspace*{\fill}
   \caption{Effects of earth matter on oscillation patterns given by
$P_{\bar{\mu}e }$.
Oscillation without matter effect in solid black, 
with matter effect of earth-model made of pure SiO$_2$ in dashed red,
and its electron number density 20\% made larger in dash-dotted blue.
$\delta = \pi/4$ and energy combination $(E_{\bar{\nu}_{\mu}} , E_{\nu_e}) =$ (500 MeV, 50MeV)
for $\bar{\mu} e$ events.
}
   \label {earth matter}
 \end{center} 
\end{figure*}

Result of numerical computations is illustrated in Fig(\ref{earth matter}),
which indicates a great sensitivity of coherent pair beam experiments
to the earth matter effect.
One might say that the matter effect contaminates
CPV effects, and it would be wise to
conduct oscillation experiments in detectors
placed on earth.
In order to avoid the flux reduction caused by emission
angles away from the beam tangential,
 a useful site distance  is limited to order 50 km.

The great sensitivity to the earth matter density
of oscillation patterns is an obstacle against
a clean measurement of CPV parameter,
but it might open a possibility of
devising a method of earth tomography
by means of the coherent neutrino pair beam,
which we hope to discuss in future.

\vspace{0.5cm}
In summary,
we showed that coherent neutrino pair beam
can provide an excellent chance of measuring
CPV parameter and distinction of the mass hierarchical patterns
if double weak events are detected.

\vspace{0.5cm}
 {\bf Acknowledgements}

We should like to thank 
M. Tanaka and K. Tsumura  for a valuable discussion.

\end{document}